\documentclass[epj]{svjour}
\usepackage{etoolbox}
\makeatletter
\providecommand{\sf@counterlist}{} 
\makeatother
\usepackage{graphicx}
\usepackage{etoolbox}
\usepackage{mathrsfs}
\usepackage{epstopdf}
\usepackage{subcaption}
\usepackage{booktabs}   
\usepackage{multirow}   
\usepackage{amsmath}    
\usepackage{subfig}
\usepackage{float}
\usepackage{amssymb}
\usepackage[title]{appendix}%
\usepackage[colorlinks=true, linkcolor=blue, citecolor=blue, urlcolor=blue]{hyperref}
\begin{document}

\title{Curled orbit and epicyclic oscillation of charged particles around the weakly magnetized black hole in the presence of Lorentz violation}

\author{Hai-Yang Zhang\inst{1} \thanks{email:hiyzhang@nuaa.edu.cn} \and Ya-Peng Hu\inst{1,2} \thanks{email:huyp@nuaa.edu.cn} \and Yu-Sen An\inst{1,2} \thanks{email:anyusen@nuaa.edu.cn(Corresponding author)}}
%
%
\institute{College of Physics, Nanjing University of Aeronautics and Astronautics, Nanjing, 210016, China \and MIIT Key Laboratory of Aerospace Information Materials and Physics,  Nanjing University of Aeronautics and Astronautics, Nanjing, 210016, China } 
\date{Received: date / Revised version: date}
%
\abstract{
In this paper, we investigate the motion of charged particles around the weakly magnetized Schwarzschild-like bumblebee black hole which has Lorentz symmetry breaking. Charged particles have curled orbits around the black hole which can only appear in the presence of external magnetic field. We investigate the effect of Lorentz violation factor on the curled orbit for both the case with and without cosmological constant. Furthermore, we investigate the harmonic oscillation behavior of the charged particles around the stable circular orbit. By using the epicyclic resonance model, we relate the harmonic oscillations of charged particles to the twin high frequency quasi-periodic oscillations observed in micro-quasars. Based on the observations of quasi-periodic oscillation, we provide a stringent constraint on the Lorentz violating parameters by using Markov Chain Monte Carlo algorithm. As the black hole shadow for Schwarzschild-like bumblebee black hole degenerates to the ordinary Schwarzschild black hole, the constraints we obtained from the quasi-period oscillation is crucial for further searching for the imprint of Lorentz symmetry breaking in our universe.
} 
\maketitle
\section{Introduction}
Black holes, as one of the most profound predictions of general relativity (GR)~\cite{wald2010general}, serve as an important laboratory for testing gravitational theories in strong gravity regime. With the direct detections of gravitational waves emitted by binary black hole mergers~\cite{LIGOScientific:2016aoc} and the release of images of supermassive central black holes in Sgr~A* and M87*~\cite{EventHorizonTelescope:2022wkp,EventHorizonTelescope:2022apq,EventHorizonTelescope:2019dse,EventHorizonTelescope:2019ths}, the black holes are not only a theoretical prediction but also a real astrophysical objects in our universe. Thus these achievements bring a lot of attentions from both theoretical side and observational side. 

Among the observations related to the supermassive celestial objects (e.g. black hole), notably, some galactic micro-quasars (central black hole surrounded by the accretion disk) \cite{Ozel:2010bz,Strohmayer:2001yn,Remillard:2002cy,Shafee:2005ef,Remillard:2006fc} receive much research interest because of the existence of high frequency quasi-periodic oscillations (HF QPO). HF QPOs, which are observed as twin peaks in X-ray power density spectra of galactic microquasars such as GRS 1915+105, XTE J1550-564, GRO J1655-40~\cite{Strohmayer:2001yn,Remillard:2002cy,Shafee:2005ef,Remillard:2006fc}, always present characteristic frequency ratios such as 3:2. The characteristic 3:2 frequency ratio of these oscillations indicates that there are resonance phenomenon in oscillations of accreting matter ~\cite{Abramowicz:2001bi,Lee:2004bp,Stuchlik:2007xt,Slany:2008pm}. In explaining the origin of this signal, strong gravity effects must be relevant, thus various explanations have been proposed by identifying the observed QPO frequencies as the resonant harmonic oscillation frequencies of test particles orbiting the central celestial objects \cite{Stella:1998mq,Kotrlova:2008xs,Stuchlik:2008fy,Kolos:2015iva}. As QPO is strongly related to the particle motion around the celestial objects, it will strongly depend on the properties of central compact celestial object. This feature makes QPO an important probe to distinguish gravitational theories and seek for the exotic objects. 
This particle (geodesic) oscillation framework has been extensively applied to diverse physical setups in recent years. Related investigations include charged black holes~\cite{Rayimbaev:2020hjs,Vrba:2020ijv,Banerjee:2021aln,Shaymatov:2021qvt,Chakraborty:2017nfu}, rotating compact objects~\cite{Banerjee:2022ffu,Boshkayev:2023rhr}, black holes in modified gravity \cite{Davlataliev:2024wdd,Davlataliev:2024yle,Shaymatov:2020yte,Guo:2025zca}, black holes surrounded by dark matter and dark energy~\cite{Stuchlik:2022xtq,Stuchlik:2021gwg,Rayimbaev:2021kjs,Xamidov:2025hrj}, and wormhole spacetime \cite{Stuchlik:2021guq,Turimov:2022iff,Lu:2024nev}. Such studies impose stringent constraints on these gravitational configurations in strong field regime and establish vital benchmarks for the theoretical viability. 

Regarding gravitational theories beyond GR, the bumblebee model stands out for its incorporation of spontaneous Lorentz symmetry breaking (LSB)~\cite{Kostelecky:1988zi,Kostelecky:1994rn,Colladay:1998fq}. Based on the standard model extension (SME) framework~\cite{Kostelecky:2004hg}, this theory introduces a non-minimally coupled vector field $B_\mu$ that acquires a nonzero vacuum expectation value by spontaneously symmetry breaking mechanism. The Lorentz symmetry breaking is induced by the nonzero vacuum expectation value of the vector field. The simplest static, spherically symmetric solution in bumblebee model has been firstly constructed in Ref.\cite{Casana:2017jkc} and then generalized to the case with cosmological constant in Ref.\cite{PhysRevD.103.044002}.  The solution obtained in \cite{Casana:2017jkc} is a slightly modified Schwarzschild metric with only radial component $g_{rr}$ shifted by the LSB parameter $\ell$. Besides above well-known approach, there are also many other approaches to incorporate Lorentz violation effect in black hole solution, the reader can consult Ref.\cite{Yang:2023wtu,Liang:2022gdk,Filho:2022yrk,Xu:2022frb} for more information about this point. While Lorentz symmetry violation is constrained to be very small in the solar system, the constraints in strong gravity regime are still needed. Thus it is still interesting to investigate the Lorentz symmetry violation phenomenon in the strong gravity scenario. Based on various models mentioned above, effect of the Lorentz breaking factor on observations have been investigated through studies of black hole shadows~\cite{Chen:2020qyp,Jha:2020pvk,Wang:2021irh,Pantig:2024kqy}, gravitational lensing~\cite{Ovgun:2018ran,DCarvalho:2021zpf,Gao:2024ejs}, quasi-normal modes~\cite{Oliveira:2018oha,Kanzi:2021cbg,Oliveira:2021abg,Gogoi:2022wyv,Guo:2023nkd,Singh:2024nvx} and thermodynamics\cite{Gomes:2018oyd,Kanzi:2019gtu,Mai:2023ggs,An:2024fzf,EslamPanah:2025zcm}. However, for the simplest bumblebee black hole raised in \cite{Casana:2017jkc}, due to  degeneracies in the $g_{tt}$ and $g_{\phi\phi}$ metric components with Schwarzschild black hole, the black hole shadow does not depend on Lorentz violating factor which makes the constraints on $\ell$ more difficult \cite{Vagnozzi:2022moj}. Constraining the parameter in this solution from other high-precision astrophysical data (such as HF QPO) constitutes part of motivation of this work.

As black hole candidate is surrounded by the accretion disk which is composed of charged particles, it is natural to take into account the effect of magnetic field. Moreover, several observations \cite{Eatough:2013nva,Will:2023nlt} further confirm that the existence of magnetic field near the black hole is ubiquitous. The magnetic field strength is weak enough in the sense that it can not back-react on the background geometry, but it is strong enough to change the particle motion around the black hole \cite{Wald:1974np,Frolov:2010mi,Kolos:2017ojf}. Although the magnetic field may be complicated near the field source, it can be approximately homogeneous at large distance scale \cite{Wald:1974np}. 
The introduction of external magnetic fields further enriches the behaviors of particle motions. 
For charged particles, the combination of gravity and Lorentz force lead to curled trajectories which only appear when magnetic field is taken into account ~\cite{Frolov:2010mi}. Moreover, the introduction of magnetic field is also helpful to explain the observational data as discussed in \cite{Kolos:2015iva,Kolos:2017ojf} . 

Testing Lorentz violating black hole background using the quasi-periodic oscillation is indeed an interesting research direction. Several works have been done by this motivation \cite{Wang:2021gtd,Mustafa:2024mvx,Jumaniyozov:2024eah,Jumaniyozov:2025wcs,Jha:2024fnh,Gu:2022grg}  \footnote{Black holes with vanishingly small spin may indeed exist in the universe due to the existence of primordial black holes which formed by other novel mechanism instead of collapse of stars.}. However, among these works, the original static spherically symmetric black holes in metric bumblebee theory \cite{Casana:2017jkc,PhysRevD.103.044002} does not receive much attention. As the black hole shadow fails to distinguish solution in Ref.\cite{Casana:2017jkc} from Schwarzschild solutions, it is necessary to focus on dynamical phenomena, such as HF QPOs, to probe the $g_{rr}$ component of the metric where LSB factor appears. Moreover, the magnetic field is not only ubiquitous in our universe but also necessary to explain the observed twin HF QPO signals. However, previous studies \cite{Wang:2021gtd,Mustafa:2024mvx,Jumaniyozov:2024eah,Jumaniyozov:2025wcs,Jha:2024fnh,Gu:2022grg} neglected the synergy effect of the magnetic field. Therefore, based on above reasons, we will investigate the effect of Lorentz violation factor $\ell$ on the charged particle motion for model \cite{Casana:2017jkc,PhysRevD.103.044002} in the presence of magnetic field. 
We show that Lorentz symmetry breaking factor $\ell$ can both significantly affect the curled motion and harmonic oscillation frequencies of charged particles around black hole. Furthermore, by using the observational data in slowly rotating micro-quasar XTE
J1550-564, we give the constraint on the Lorentz symmetry breaking factor $\ell$ by using the Markov Chain Monte Carlo algorithm. Our results would help to bridge the connection between astrophysical observations and fundamental theories about Lorentz symmetry breaking.

This paper is organized as follows. In Sec.\ref{bumblebee}, we have briefly reviewed the bumblebee gravity model and the black hole solution predicted by this theory. In Sec.\ref{motion}, we investigate the trajectory of the charged particle immersed in an uniform magnetic field with different LSB parameter. We study the effect caused by LSB parameter both in the case with and without cosmological constant. In Sec.\ref{QPO}, we investigate the harmonic oscillations of charged particles around the stable circular orbit. We compute the radial and latitudinal epicyclic frequencies $\nu_{r},\nu_\theta$ in magnetized Schwarzschild-like black hole and further explore the effect of the cosmological constant on these epicyclic oscillation frequencies. In Sec.\ref{mcmc}, by using resonant model and directly identifying the epicyclic oscillation frequencies as HF QPO frequencies observed in microquasar XTE J1550-564, we provide a plausible constraints on the value of Lorentz symmetry breaking factor $\ell$ by using MCMC method. In Sec. \ref{conclusion}, we summarize our main results and give a vision for the future work. Throughout the paper, we use the system of geometric units in which $G=1=c$. We will restore the dimensionality when comparing with the observation. 
\section{Review of Schwarzschild-like bumblebee black hole }\label{bumblebee}
The search for Lorentz symmetry breaking, as a potential remnant of quantum gravity at low energies, has motivated extensive studies in modified gravitational frameworks. Among these, bumblebee gravity emerges as a compelling model, where spontaneous Lorentz symmetry breaking is triggered by the nonzero vacuum expectation value (VEV) of the vector field $B_\mu$. 

We briefly review details of bumblebee models in this section. The action of bumblebee model containing the cosmological constant $\Lambda$ was given by \cite{PhysRevD.103.044002}
\begin{equation}\begin{aligned}
S_{B}=\int d^{4}x\sqrt{-g}\left[\frac{1}{2\kappa}(R-2\Lambda)+\frac{\xi}{2\kappa}B^{\mu}B^{\nu}R_{\mu\nu}\right. \\ \left.-\frac{1}{4}B_{\mu\nu}B^{\mu\nu}-V(B^{\mu}B_{\mu}\pm b^{2})+\mathcal{L}_{M}\right],
\end{aligned}\end{equation}
where $\kappa=\frac{8\pi G}{c^4}$, $B_{\mu\nu}\equiv\partial_{\mu}B_{\nu}-\partial_{\nu}B_{\mu}$. In the action, $\xi$ is the coupling constant between bumblebee field and Ricci curvature and $b^2$ is a positive real constant, $\mathcal{L}_M$ is the Lagrangian of additional matter field that may appear in the theory. For more discussions on the action, the reader can consult Ref. \cite{Casana:2017jkc,PhysRevD.74.045001,PhysRevD.88.025005}. 

The minimum of the potential $V$ provides a non-zero VEV of bumblebee field and thus controls the spontaneous Lorentz symmetry breaking. For non-negative polynomial type potential $V(X)$, the VEV of bumblebee is determined by 
\begin{equation}
    V(B^{\mu}B_{\mu}\pm b^{2})=0
\end{equation}
which is equivalent to $B^\mu B_\mu=\mp b^2$, where $\pm$ sign determines whether the VEV $b_\mu$ is space-like or time-like. When $B^{\mu}$ settles to its VEV $b^{\mu}$, the Lorentz symmetry is spontaneously broken.

One can first consider the case without cosmological constant $\Lambda = 0$ and seek for the simplest spherically symmetric vacuum solution . The simplest ansatz is to choose the space-like bumblebee vector field $B_\mu$ which can be written as
\begin{equation}
    B_\mu=b_\mu=(0,b_r(r),0,0)
    \label{ansatz1}
\end{equation}
where the field strength vanishes $b_{\mu\nu}=\partial_{\mu}b_{\nu}-\partial_{\nu}b_{\mu}=0$.
Moreover, for the static spherically symmetric spacetime, the metric ansatz reads:
\begin{equation}
ds^2 = -h(r)dt^2 + \frac{1}{f(r)}dr^2 + r^2 (d\theta^2 + \sin^2\theta d\phi^2).
\label{ansatz2}
\end{equation}


In Ref.\cite{Casana:2017jkc}, minimum potential condition $V'=0$ is imposed to further simplify the equation of motion. The simplified equations of motion in the presence of the bumblebee field read
\begin{equation}
\begin{aligned}
R_{\mu\nu}+\gamma b_{\mu}b^{\alpha}R_{\alpha\nu}+\gamma b_{\nu}b^{\alpha}R_{\alpha\mu}-\frac{\gamma}{2}b^{\alpha}b^{\beta}R_{\alpha\beta}g_{\mu\nu}- \\
\frac{\gamma}{2}\nabla_{\alpha}\nabla_{\mu}(b^{\alpha}b_{\nu})-\frac{\gamma}{2}\nabla_{\alpha}\nabla_{\nu}(b^{\alpha}b_{\mu})+\frac{\gamma}{2}\nabla^{2}(b_{\mu}b_{\nu})=\mathrm{0},
\label{motion equation of bumblebee 1}
\end{aligned}
\end{equation}
\begin{equation}
\nabla^{\mu}b_{\mu\nu}=-\frac{\gamma}{\kappa}b^{\mu}R_{\mu\nu}.
\label{motion equation of bumblebee 2}
\end{equation}

Substituting into the ansatz (\ref{ansatz1}),(\ref{ansatz2}), the equation of bumblebee field is trivially satisfied and the form of extended Einstein equation is
\begin{equation}(1+\frac{\ell}{2})R_{tt}+\frac{\ell}{2r}[\partial_{r}h(r)f(r)-\partial_{r}f(r)h(r)]=0,
\label{Einstein equation 1}
\end{equation}
\begin{equation}
(1+\frac{3\ell}{2})R_{rr}=0,
\label{Einstein equation 2}
\end{equation}
\begin{equation}
(1+\ell)R_{\theta\theta}-\ell[\frac{1}{2}r^{2}f(r)R_{rr}+1]=0.
\label{Einstein equation 3}
\end{equation}
where $\ell$ is a constant which is related to vacuum value of bumblebee field as $\ell=\xi b^{2}$. 
Solving these coupled equations yields a Schwarzschild-like metric, 
\begin{equation}
\begin{aligned}
ds^{2}&=-\left(1-\frac{2M}r\right)dt^2+(1+\ell)\left(1-\frac{2M}r\right)^{-1}dr^2\\&+r^2(d\theta^2+\sin^2\theta d\phi^2).
\label{metric of bumblebee}
\end{aligned}
\end{equation}
where $M$ is the mass parameter of black hole and it can be easily seen that the constant $\ell$ measures the magnitude of Lorentz symmetry breaking which is called Lorentz-violating factor. In the limit $\ell \to 0 $ we recover the usual Schwarzschild metric. By computing the Kretschmann scalar
\begin{equation}
R_{\mu\nu\alpha\beta}R^{\mu\nu\alpha\beta}=\frac{4(12M^2+4\ell Mr+\ell^2r^2)}{r^6(\ell+1)^2}.
\label{Kretschmann}
\end{equation}
we find that it is different from the case for Schwarzschild black hole, thus we can tell that this is indeed a distinct black hole metric.

In Ref.\cite{PhysRevD.103.044002}, a further generalization has been made to find the bumblebee black hole solution with cosmological constant. The procedure to find the solution is similar to the case without cosmological constant despite the linear potential of the bumblebee field should be imposed
\begin{equation}
    V =V(X)= \frac{\lambda}{2} \left( B^\mu B_\mu - b^2 \right) 
    \label{linear potential 1}
\end{equation}
where $X=B^\mu B_\mu-b^2$ and $\lambda$ is the Lagrangian multiplier. For this potential form, the condition $V'=0$ should not be satisfied \footnote{The condition $B^{\mu}B_{\mu}=b^{2}$ is still satisfied in this case as a result of the equation of Lagrangian multiplier $\lambda$.}
\begin{equation}
    V'(X) = \frac{\lambda}{2}.
    \label{linear potential 2}
\end{equation}
This is a crucial difference from the case without cosmological constant. The static, spherical symmetric black hole metric with the inclusion of cosmological constant reads
\begin{equation}\begin{gathered}
ds^2=-\left[1-\frac{2M}{r}-(1+\ell)\frac{\Lambda_e}{3}r^2\right]dt^2\\
+(1+\ell)\left[1-\frac{2M}{r}-(1+\ell)\frac{\Lambda_e}{3}r^2\right]^{-1}dr^2\\+r^2\left(d\theta^2+\sin^2\theta d\phi^2\right),
\label{metric Lambda_e}
\end{gathered}\end{equation}
here, $\Lambda_{e}=\frac{\kappa\lambda}{\xi}$ is an effective cosmological constant and the cosmological constant $\Lambda$ must satisfy a strong constraint $\Lambda=\frac{\kappa\lambda}{\xi}(1+\ell)$. When $\Lambda_e=0$, the line element will return to the solution in \cite{Casana:2017jkc}. Depending on the sign of $\Lambda_e$, the solutions bifurcate into two classes:
\begin{itemize}
    \item \textbf{Schwarzschild-anti-de Sitter-like black holes} 
    
    ($\Lambda_e < 0$): Characterized by a single event horizon at radius $r_+$, which diminishes as $\ell$ increases.
    \item \textbf{Schwarzschild-de Sitter-like black holes} 
    
    ($\Lambda_e > 0$): Having both an event horizon $r_+$ and a cosmological horizon $r_c$ when $\Lambda_{e}$ satisfies $0 < \Lambda_e < [9M^2(1+\ell)]^{-1}$. Increasing $\ell$ enlarges $r_+$ while reducing $r_c$.
\end{itemize}

As the observations support that the cosmological constant of our universe is positive, we will only focus on the $\Lambda_{e}>0$ case in the following. Moreover, it needs to be stressed that the spacetime is neither asymptotically de Sitter nor anti-de Sitter due to the irreducible $(1+\ell)$ factor. These facts are the reflection of Lorentz symmetry violation for the underlying spacetime, which offers novel insights into quantum gravity phenomenology.
\section{Motion of the charged particle near the weakly magnetized bumblebee black hole}\label{motion}
Based on observations, it has been confirmed that the existence of magnetic field near the black hole is ubiquitous. For theoretical simplicity, we impose the uniform magnetic field perpendicular to the equatorial plane to investigate the effect of magnetic field on the motion of particles \cite{Wald:1974np}.

While it seems to be inconsistent for choosing uniform magnetic field around bumblebee black hole as it does not satisfy sourceless Maxwell equation\footnote{The uniform magnetic field satisfies the sourceless Maxwell equation only for Ricci flat spacetime, for more discussions the reader can consult Ref.\cite{Shaymatov:2020yte,Wald:1974np,Azreg-Ainou:2016tkt}.}, this point is not sufficient to support that there is no uniform magnetic field around Schwarzschild-like bumblebee black hole. Besides non-minimal coupling to Ricci curvature,  bumblebee field is also non minimally coupled to electro-magnetic field which will provide the source term and significantly modify the Maxwell equation \cite{Seifert:2009gi}. The electro-magnetic field should satisfy the sourced Maxwell equation in this case $\nabla_{\mu}F^{\mu\nu}=J^{\nu}$ instead of sourceless Maxwell equation which is different from the cases of other modified gravities \cite{Shaymatov:2020yte,Azreg-Ainou:2016tkt}. Actually, there are various distinct forms of the non-minimal coupling between bumblebee field and electromagnetic field \cite{Seifert:2009gi,Maluf:2015hda,Liu:2024axg}. Although there still remains some controversies among them, we would like to postpone the consideration of this theoretical complexity to the future study. We thus choose to impose that the magnetic field can be approximately uniform around the black hole spacetime (at least to leading order) in order to get the analytical understanding of the system.

The electro-magnetic four-vector potential $A_\mu$ corresponding to the uniform magnetic field takes the following form as shown in Ref.\cite{Kolos:2015iva,Wald:1974np}
\begin{equation}A^{\mu}=\frac{B}{2} \xi_{(\phi)}^{\mu}.
\label{four-vector potential in dicretion}
\end{equation}
By Stokes theorem, it can be directly seen that the magnetic field is uniform and oriented perpendicularly to the equatorial plane of the black hole spacetime. The non-zero magnetic field will exert Lorentz force on the charged particles moving in the spacetime, thus the dynamical equation of a charged particle is 
\begin{equation}
m\frac{\mathrm{d}u^{\mu}}{\mathrm{d}\tau}=qF_{\nu}^{\mu}u^{\nu},
\label{dynamical equation of a charged particle}
\end{equation}
where $u^{\mu}=\frac{\mathrm{d}x^{\mu}}{\mathrm{d}\tau}$ is the four-velocity of the particle, satisfying $u_\mu{u^\mu}=-1$,$\tau$ is the proper time of the particle and $F_{\mu\nu}=A_{\mu,\nu}-A_{\nu,\mu}$ is the antisymmetric tensor of the electromagnetic field.

The Hamiltonian for the charged particles can be written in the following form
\begin{equation}H=\frac{1}{2}g^{\alpha\beta}\Big(\pi_{\alpha}-qA_{\alpha}\Big)\Big(\pi_{\beta}-qA_{\beta}\Big)+\frac{1}{2} m^{2},
\label{Hamiltonian}\end{equation}
$\pi_\mu$ is the canonical four-momentum which is distinct from the mechanical four-momentum $p_\mu=mu_\mu$. The two kinds of momentum satisfy the relation $\pi_\mu=p_\mu+qA_\mu$, where $A_{\mu}$ is the covariant electromagnetic field with only nonzero component $A_{\phi}=\frac{B}{2}r^{2}\sin^{2}\theta$. The motion of charged particles satisfy the Hamiltonian canonical equation, which is 
\begin{equation}m\frac{\mathrm{d}x^{\mu}}{\mathrm{d}\tau}\equiv p^{\mu}=\frac{\partial H}{\partial\pi_{\mu}},\quad m\frac{\mathrm{d}\pi_{\mu}}{\mathrm{d}\tau}=-\frac{\partial H}{\partial x^{\mu}},
\label{Hamiltonian canonical equations}\end{equation}
As the Hamiltonian does not depend on coordinate $t$ and $\phi$ explicitly, two conserved quantities can be deduced from the Hamiltonian canonical equations which is respectively 
\begin{equation}\begin{aligned}&E=-\pi_{t}=m h\left(r\right)\frac{\mathrm{d}t}{\mathrm{d}\tau},\\&L=\pi_{\phi}=mr^{2}\sin^{2}\theta\biggl(\frac{\mathrm{d}\phi}{\mathrm{d}\tau} + \frac{qB}{2m}\biggr).\label{conserved quantities}\end{aligned}\end{equation}
For convenience, in the following analysis of particle motion, we will use the following dimensionless quantities 
\begin{equation}\mathcal{E}=\frac{E}{m},\quad\mathcal{L}=\frac{L}{m},\quad\mathcal{B}=\frac{qB}{2m}.
\label{dimensionless quantities}\end{equation}
which is called specific energy, specific angular momentum and magnetic parameter respectively. Note that instead of magnetic field $B$, the frequency $\mathcal{B}$ which we call magnetic parameter is directly related to the particle motion. In the below, we will directly use this $\mathcal{B}$ to represent the magnitude of the magnetic field. Based on observation related to particle motion, in order to know the real magnitude of magnetic field, we also need to know the kind of charged particles which determines the charge to mass ratio $q/m$. For different types of charged particles, the estimated magnetic field will be different \cite{Kolos:2015iva,Kolos:2017ojf}. 

For the massive charged particles, the geodesic must satisfy $u_\mu{u^\mu}=-1$, which can be written in terms of $\mathcal{E}$,$\mathcal{L}$,$\mathcal{B}$ as 
\begin{equation}
\begin{aligned}
\frac{1}{f(r)}(\frac{dr}{d\tau})^{2}+r^{2}(\frac{d\theta}{d\tau})^{2}&=-1+\frac{\mathcal{E}^{2}}{h(r)}-(\frac{\mathcal{L}}{r\sin\theta}-\mathcal{B} r \sin\theta)^{2} \\& =\frac{1}{h(r)}(\mathcal{E}^{2}-V_{eff}(r,\theta))
\label{equation of motion with metric}
\end{aligned}
\end{equation}
where the effective potential is defined as 
\begin{equation}
    V_{eff}(r,\theta)=h(r)(1+(\frac{\mathcal{L}}{r\sin\theta}-\mathcal{B} r \sin\theta)^{2})\label{effect}
\end{equation}
which is a 2d function in terms of variables $r$ and $\theta$. 
As the LHS of Eq.(\ref{equation of motion with metric}) is always positive, thus the effective potential is useful to bound the particle's trajectories such that particles can only move in the region $\mathcal{E}^{2}>V_{eff}(r,\theta)$. 
Stationary points of the effective potential are given by the equations 
\begin{equation}
    \partial_{r}V_{eff}(r,\theta)=0, \quad \partial_{\theta}V_{eff}(r,\theta)=0
\end{equation}
the solution of second equation is $\theta=\frac{\pi}{2}$, thus the stationary points of effective potential must lie on the equatorial plane. In this section, we first consider the charged particle motion on the equatorial plane, so we set $\theta=\pi/2$ in this section.

After setting $\theta=\frac{\pi}{2}$, the radial equation of motion is simplified as
\begin{equation}\begin{aligned}\frac{1}{f(r)}(\frac{dr}{d\tau})^{2}&=\frac{1}{h(r)}\left\{\mathcal{E}^{2}-h(r)[1+\frac{(\mathcal{L}-\mathcal{B}r^2)^{2}}{r^{2}}]\right\}\\&= \frac{1}{h(r)}\left\{\mathcal{E}^{2}-\mathcal{V}_{\mathrm{eff}}(r)\right\}.
\label{radial equation of motion with metric}\end{aligned}\end{equation}
where the radial effective potential is 
\begin{equation}
    V_{eff}(r)=h(r)[1+\frac{(\mathcal{L}-\mathcal{B}r^2)^{2}}{r^{2}}].
\label{effective potential}
\end{equation}
We can first concentrate on the circular orbit, 
the radius of the circular orbit is determined by the following condition of effective potential 
\begin{equation}
V_{eff}(r_0)=\mathcal{E}^2, \quad \frac{\partial V_{eff}(r)}{\partial r}|_{r=r_0}=0.
\label{radial effective potential}
\end{equation}
By plugging in the metric Eq.(\ref{metric Lambda_e}), the specific energy $\mathcal{E}$ and angular momentum  $\mathcal{L}$ on the circular orbit can be calculated in terms of the circular orbit radius $r_0$ as
\begin{equation}
\begin{aligned}
&\mathcal{L}=\frac{r_0^{2}\left[-3\mathcal{B}+\mathcal{B}(\ell+1) r_0^{3} \Lambda_{e}\right]}{3(r_0-3)}
\\&+\frac{r_0\sqrt{\left(-3\mathcal{B}(r_0-2) r_0+\mathcal{B}(\ell+1) r_0^{4} \Lambda_{e}\right)^{2}
-\mathcal{G}_1}}{3(r_0-3)},\\ &\mathcal{E}=\sqrt{-\frac{\left[6-3r_0+(\ell+1)r_0^3\Lambda_{e}\right] \mathcal{G}_2}{3 r_0^3}}.
\label{expressions of E and L }
\end{aligned}
\end{equation}
where 
\begin{equation*}\begin{aligned}
    \mathcal{G}_1&=3(r_0-3)\left(-3+(\ell+1) r_0^{3} \Lambda_{e}\right),
    \\
    \mathcal{G}_2&=\left[\mathrm{\mathcal{B}}^2r_0^4+(1-2\mathrm{\mathcal{B}\mathcal{L}})r_0^2+\mathcal{L}^2\right].
\end{aligned}    
\end{equation*}
Note that for simplicity, we choose to redefine the dimensionless variable $r/M \to r$ and $\Lambda_{e} M^{2} \to \Lambda_{e}$ for the computation and restore the dimension when we need to compare with observations.

Note that when $\Lambda_e=0$, the expressions of $\mathcal{E}$ and $\mathcal{L}$ reduce to the result of Schwarzschild black hole \cite{Kolos:2015iva}. This means that without cosmological constant, Lorentz violation can not influence the particle's circular orbit. This is expected since in $\Lambda_{e}=0$ case, the Lorentz violation factor $\ell$ only affects the $g_{rr}$ component, while effective potential in Eq.(\ref{effect}) does not depend on it. This is also the reason why the black hole shadow observation fails to distinguish Schwarzschild like bumblebee black hole in Eq.(\ref{metric of bumblebee}) from the ordinary Schwarzschild black hole \cite{Vagnozzi:2022moj}. For the innermost stable circular orbit(ISCO), we must have $\frac{\partial^2 V_{eff}}{\partial r^2}|_{r=r_{isco}}$ $=0$. For the same reason, without cosmological constant, the ISCO radius $r_{isco}$ of the Schwarzschild-like BH in bumblebee gravity does not depend on the Lorentz symmetry breaking parameter $\ell$. 

Although the Lorentz violation factor $\ell$ has relatively small effect on the circular orbit, it will strongly influence the behavior of other kinds of orbit. This point can be seen in the following way, by considering the perturbation $r+\delta r$ of Eq.(\ref{radial equation of motion with metric}) near the circular orbit, the perturbation equation to the leading order becomes 
\begin{equation}
\frac{d^{2}\delta r}{d\tau^{2}}+\frac{V_{eff}''(r_{0})}{2(1+l)}\delta r=0    
\end{equation}
thus although effective potential for metric Eq.(\ref{metric of bumblebee})  is the same as the case in Schwarzschild metric, the Lorentz violation factor $\ell$ will shift the oscillation frequency of perturbation. 

\begin{figure*}[h!]
    \centering
    \includegraphics[width=1\linewidth]{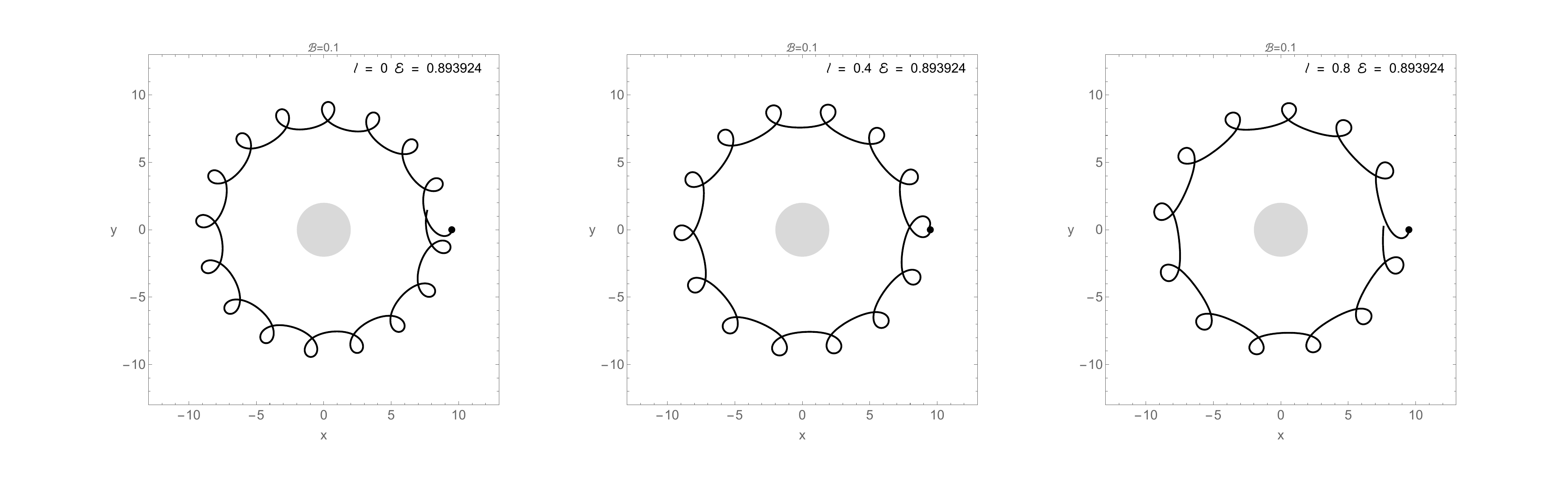}
    \caption{Trajectories of the particle at the equatorial plane ($\theta=\frac{\pi}{2},\dot{\theta}=0$) around the magnetized bumblebee black hole for magnetic field parameter $\mathcal{B}=0.1$, effective cosmological constant $\Lambda_e=0$ and LSB parameter $\ell=0, 0.4, 0.8$. The motion of the particle is plotted from a given initial radius $r_0=9.5,\mathcal{L}=7.96$. }
    \label{fig1}
\end{figure*}

So below, we will go beyond the circular orbit and investigate other kinds of bounded orbit for bumblebee black hole. 
Solving the coupled differential equations in $\phi$ direction and $r$ direction 
\begin{equation} 
\dot{\phi}=\frac{\mathcal{L}}{r^2}-\mathcal{B},
\label{ecc}
\end{equation}
\begin{equation}
\dot{r}^2=\frac{\mathcal{E}^2}{1+\ell}-\frac{\left[1-\frac{2}{r}-(1+\ell)\frac{\Lambda_e}{3}r^2\right]}{1+\ell}\left[1+r^2(\frac{\mathcal{L}}{r^2}-\mathcal{B})^2\right]
\label{ecc1}
\end{equation}
at fixed parameters $\mathcal{L}$, $\mathcal{B}$, $\mathcal{E}$ and $\Lambda_{e}$ will give us the general trajectories of charged particles on the equatorial plane. Before numerically solving this equation, we can firstly find that for the equation of $\phi$, $\dot{\phi}$ can be either negative or positive depending on the value of $r$ when $\mathcal{L}$ and $\mathcal{B}$ have the same sign. This means that $\phi$ may not be monotonic for some choice of parameters. This phenomenon, which is called curled motion in Ref.\cite{Frolov:2010mi}, is distinct from the case without magnetic field where $\phi$ is a monotonic function. As the motion with non-monotonic $\phi$ is the most novel feature caused by magnetic field, we will consider this kind of orbit in bumblebee black hole. 
We plot the curled motion for the case $\Lambda_{e}=0$ in Fig.\ref{fig1}, we find that number of curls for one cycle decreases as $\ell$ growing. The number of curls is the times the particle oscillates in radial direction when particles move around the black hole for one time. Thus it is the ratio between the frequency in $r$ direction and $\phi$ direction
\begin{equation}
   N=\frac{\omega_{r}}{\omega_{\phi}}
\end{equation}
where $\omega_{\phi}=\dot{\phi}/2\pi$. From Eq.(\ref{ecc}), the angular velocity (frequency) in $\phi$ direction does not depend on Lorentz violation factor $\ell$, while the frequencies for $r$ direction does depend on it. Because $\omega_{r}^{2}=\frac{V_{eff}''(r_{0})}{2(1+l)}$, thus the curled number will decrease in terms of $(1+l)^{-1/2}$ when increasing Lorentz violating factor $\ell$. Our numerical result fits this expectation well. 

\begin{figure*}[h!]
    \centering
    \includegraphics[width=0.945\linewidth]{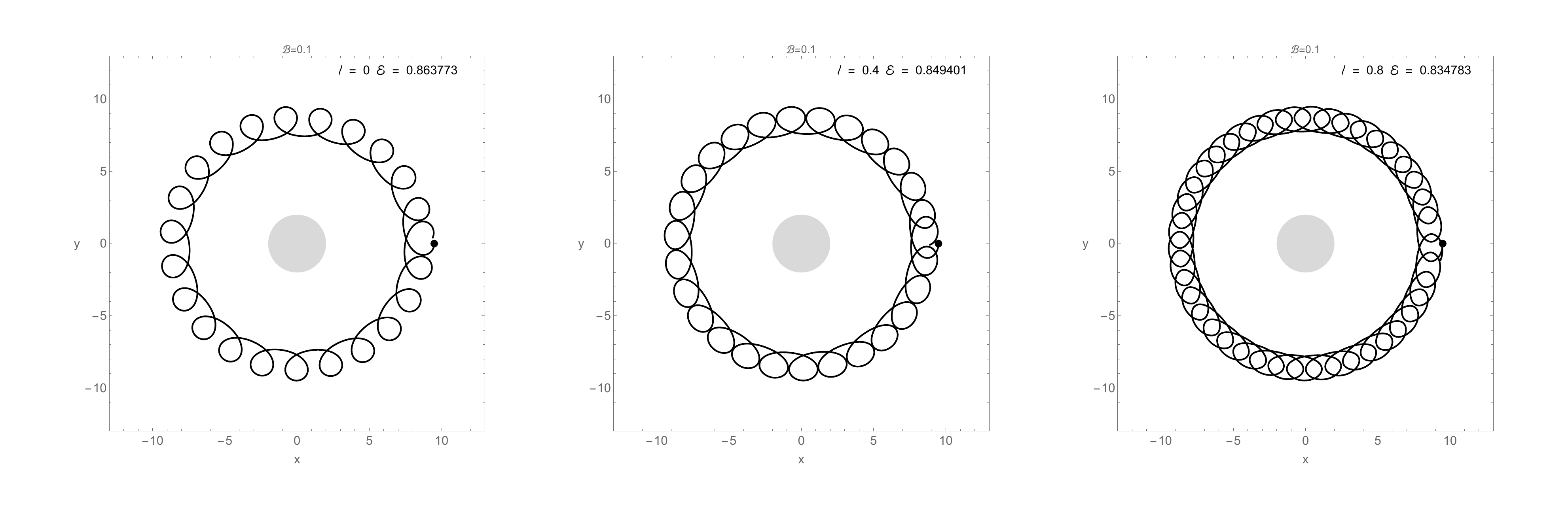}
    \caption{Trajectories of the particle at the equatorial plane ($\theta=\frac{\pi}{2},\dot{\theta}=0$) around the magnetized bumblebee black hole for magnetic field parameter $\mathcal{B}=0.1$ in the presence of positive cosmological constant. The effective cosmological constant is chosen to be $\Lambda_e=0.002$ for illustration and the motion of  particle is plotted from a given initial radius $r_0=9.5,\mathcal{L}=7.57$. When increasing LSB parameter $\ell$, the number of curls will increase. }
    \label{fig2}
\end{figure*}

For the case with cosmological constant, things will be different. With cosmological constant, the effective potential will be changed as the Lorentz violating factor $\ell$ affect the metric component $g_{tt}$. Thus when $\ell$ increases, the term $V''_{eff}(r_{0})$ in $\omega_r$ is also significantly changed by both changes of circular orbit radius and the form of effective potential which makes the change rule of curls number sophisticated. We numerically plot the curled trajectories in Fig.\ref{fig2} for the case with cosmological constant $\Lambda_{e}=0.002$ , we find that for this case, increasing the Lorentz violation factor $\ell$ will increase the curled number which is contrary to the $\Lambda_{e}=0$ case. 

\section{Harmonic oscillations and twin HF QPOs in the microquasars }\label{QPO}
From the above analysis, we conclude that the Lorentz violation factor indeed affect the oscillation frequencies of curled orbit. Thus it will be interesting to see if this influence can be reflected in observations. In this section, by invoking the resonance model, we investigate that whether the harmonic oscillations of charged particles around the stable circular orbit can be related to the observation of HF QPOs in microquasars.   

As Ref.\cite{Ingram:2019mna} mentioned, HF QPO is generally thought to be caused by the resonance phenomenon of charged particle oscillations in accretion disk as there are rational ratios between the frequencies
\begin{equation}
    \nu_{U}:\nu_{L}=3:2.
\end{equation}
In this work, we still follow the resonance approach in Ref.\cite{Kolos:2015iva} which directly identifies the red-shifted frequencies of radial and latitude harmonic oscillations with $\nu_{U}$ and $\nu_{L}$ in HF QPO observations. Thus the crucial thing is to check that if the red-shifted oscillation frequencies can exhibit $3:2$ or $2:3$ ratios. 

We will first calculate the red shifted harmonic oscillation frequencies. Different from the above section, we will also consider the off equatorial plane oscillation in this section. We consider the small perturbations in radial $(\{ r\rightarrow r_0+\delta r,\theta_{0}=\frac{\pi}{2}\})$ and latitudinal $(\{r=r_{0},\theta\rightarrow \frac{\pi}{2}+\delta\theta\})$ directions from the stable circular orbit. 
By expanding Eq.(\ref{equation of motion with metric}), the equation of perturbation $\delta r$ and $\delta \theta$ can be written as the form of harmonic oscillations  \cite{Kolos:2015iva,Abramowicz:2004tm}
\begin{equation}
\frac{d^{2}\delta r}{d\tau^{2}}+\omega_r^2\delta r=0,\quad 
\frac{d^{2}\delta \theta}{d\tau^{2}}+\omega_\theta^2\delta\theta=0.
\label{oscillation equations}\end{equation}
Here, the epicyclic frequencies are the values that are measured by local observers of the oscillating particles which read 
\begin{equation}\begin{aligned}
     &\omega_{r}^{2}=\frac{f(r_{0})}{2 h(r_0)}\partial_{r}^{2}V_{eff}|_{r=r_{0},\theta=\pi/2},
     \\&
    \omega_{\theta}^{2}=\frac{1}{2r_{0}^{2}h(r_{0})}\partial_{\theta}^{2}V_{eff}|_{r=r_{0},\theta=\pi/2}
\end{aligned}  \end{equation}

The observed frequencies are measured by the observers at infinity.  The equation of motion (\ref{oscillation equations}) at infinity can be re-written as
\begin{equation}\frac{d^2\delta r}{dt^2}+\Omega_r^2\delta r=0 ,\quad\frac{d^2\delta\theta}{dt^2}+\Omega_\theta^2\delta\theta=0 \end{equation}
where the frequencies are red-shifted as 
\begin{equation}
\Omega_r^2 =\frac{\omega^2_r}{\dot{t}^2}, \quad 
\Omega_\theta^2 =\frac{\omega^2_\theta}{\dot{t}^2}.  
\label{frenquency from infinity}
\end{equation}
The red-shift factor can be calculated in terms of the relation between $t$ and $\tau$ in Eq.(\ref{conserved quantities}), where $\mathcal{E}=\sqrt{V_{eff}(r_{0},\frac{\pi}{2})}$ as the harmonic oscillations are around the circular orbit. 
Plugging in the bumblebee Schwarzschild-like metric in Eq (\ref{metric of bumblebee}) and the expressions of two conserved quantities in Eq (\ref{conserved quantities}), we can get the expressions of $\Omega_r^2$ and $\Omega_\theta^2$
\begin{equation}
   \Omega_r^2 =\frac{r_0 - 6 + 4\mathcal{B}^2 r_0^3 \, \mathcal{H} + 8\mathcal{B}^3 (r_0 - 2) r_0^3 \mathcal{F}}{r_0^4 (\ell +1) \left(1+4 r_0^2 \mathcal{B}^2\right)},
\label{frequency of r direction}
\end{equation}
\begin{equation}\Omega_\theta^2 =\frac{1}{r_0^3},
\label{frequency of theta direction}\end{equation}
where 
\begin{equation*}
    \mathcal{F}=\sqrt{\mathcal{B}^2 r_0^2 (r_0 - 2)^2 + r_0 - 3},
\end{equation*}
\begin{equation*}
    \mathcal{H} = 2\mathcal{B}^2 r_0^3 - 8\mathcal{B}^2 r_0^2 + 8\mathcal{B}^2 r_0 + r_0 - 4.
\end{equation*}
 
Interestingly, the latitude frequency does not depend on the magnetic field and the metric. 
For vanishing magnetic parameter $\mathcal{B}=0$, the radial frequency reduces to the following simple expression
\begin{equation}\Omega_r^2 =\frac{r_0-6}{(1+\ell)r_0^4}.
\label{relation of vupp-M}\end{equation}

In the situation with cosmological constant $\Lambda_e$, the expressions of $\Omega_r^2$ and $\Omega_\theta^2$ are as follows
\begin{equation}\begin{aligned}
\Omega_r^2& = \frac{r_0^3 \left(\mathcal{J}_2-\mathcal{J}_1\right)+9 r_0-54}{9r_0^4 (\ell +1) \left(4 r_0^2 \mathcal{B}^2+1\right)},
\\&\Omega_\theta^2 =\frac{1}{r_0^{3}}-\frac{1}{3}(\ell+1) \Lambda_e,
\label{fds}
\end{aligned}\end{equation}
where
\begin{equation*}
       \mathcal{J}_1=8 \mathcal{B}^3 \left(r_0^3 (\ell +1) \Lambda _e-3 r_0+6\right) \mathcal{J}_3+3 \left(4 r_0-15\right) (\ell +1) \Lambda _e,
\end{equation*}
\begin{equation*}\begin{aligned}
     \mathcal{J}_2=&12 \mathcal{B}^2 \left( 3r_0-r_0^2 \left(5 r_0-18\right) (\ell +1) \Lambda _e-12\right)\\&+8 r_0 \mathcal{B}^4 \left(r_0^3 (\ell +1) \Lambda _e-3 r_0+6\right){}^2,
\end{aligned}
\end{equation*}
\begin{equation*}\begin{aligned}
     \mathcal{J}_{3}^2=&r_0^8 \mathcal{B}^2 (\ell +1)^2 \Lambda _e^2+9 r_0 \left(r_0 \left(r_0-2\right){}^2 \mathcal{B}^2+1\right)-27\\&-3 r_0^3 (\ell +1) \Lambda _e \left(2 \left(r_0-2\right) r_0^2 \mathcal{B}^2+r_0-3\right).
\end{aligned}
\end{equation*}
When the magnetic parameter $\mathcal{B}$ is zero, the radial frequency is given by the simplified expression below
\begin{equation}
\Omega_r^2 =\frac{r_0-6}{(1+\ell) r_0^{4}}+\left(\frac{5}{r_0}-\frac{4}{3}\right) \Lambda_e.
\end{equation}

In astronomy, from the power density spectrum obtained by fast Fourier transform of the light curve, the twin HF QPOs can be seen on the spectrum. 
To compare with the observation , we need to restore the dimensionality of epicyclic frequencies that are given by 
\begin{equation}\nu_\beta=\frac{1}{2\pi}\frac{c^3}{GM} \Omega_{\beta}[{\rm Hz}]
\label{frequency in astronomy}\end{equation}
where $\beta=r,\theta$.

\begin{figure*}[ht!]
    \centering
    \includegraphics[width=1.0\linewidth]{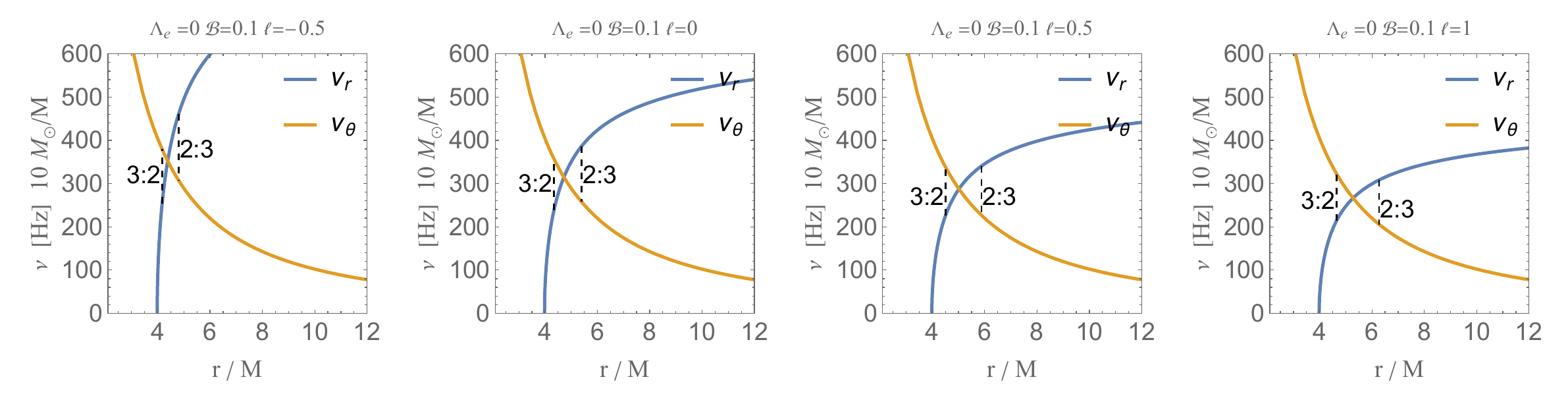}
    \includegraphics[width=1.0\linewidth]{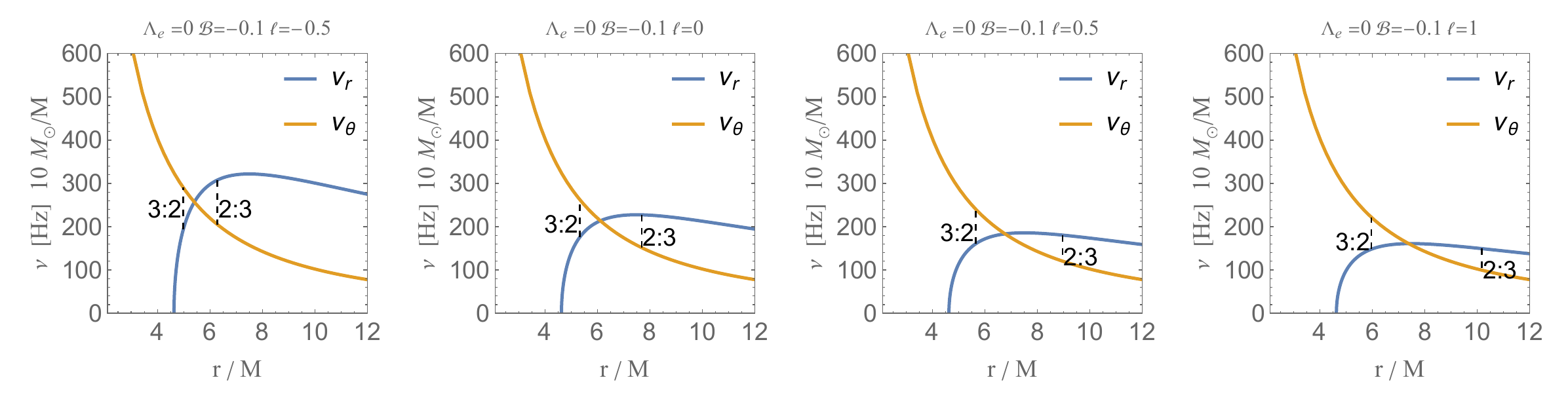}
    \caption{The epicyclic frequencies of the charged particle around the Schwarzschild-like black hole in the bumblebee gravity immersed in an external uniform magnetic field. LSB parameter $\ell$ and  magnetic parameter $\mathcal{B}$ are chosen to be $\ell=-0.5,0,0.5,1$, $\mathcal{B}=\pm 0.1$ respectively for illustration. The black dashed line represents the position where the ratio of $\nu_\theta$ and $\nu_r$ is 3:2 and 2:3. This figure is for the case without cosmological constant. }
    \label{fig3}
\end{figure*}

\begin{figure*}[h!]
    \centering
    \includegraphics[width=1.0\linewidth]{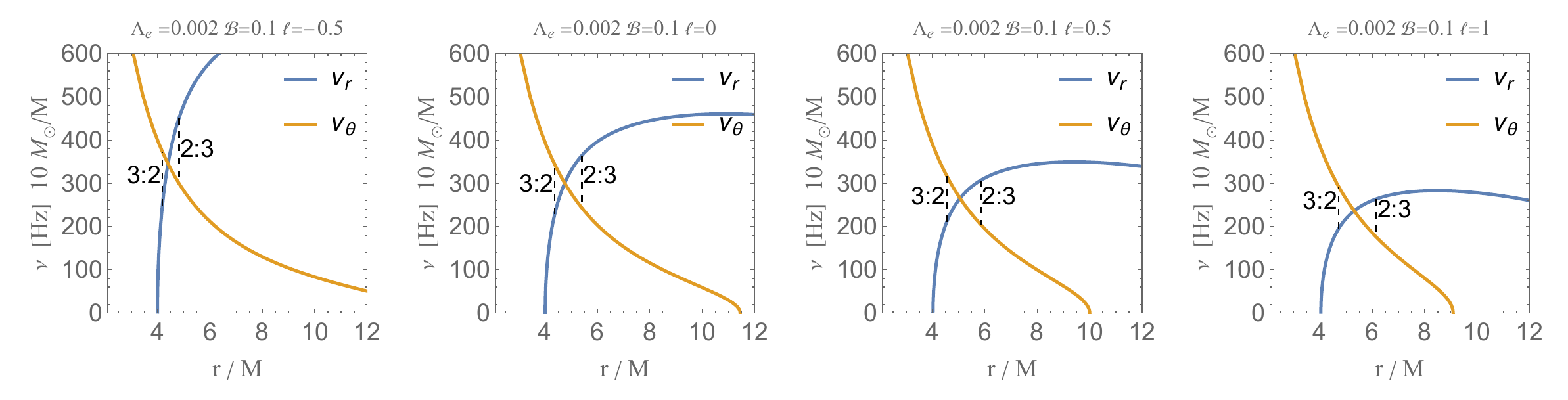}
    \includegraphics[width=1.0\linewidth]{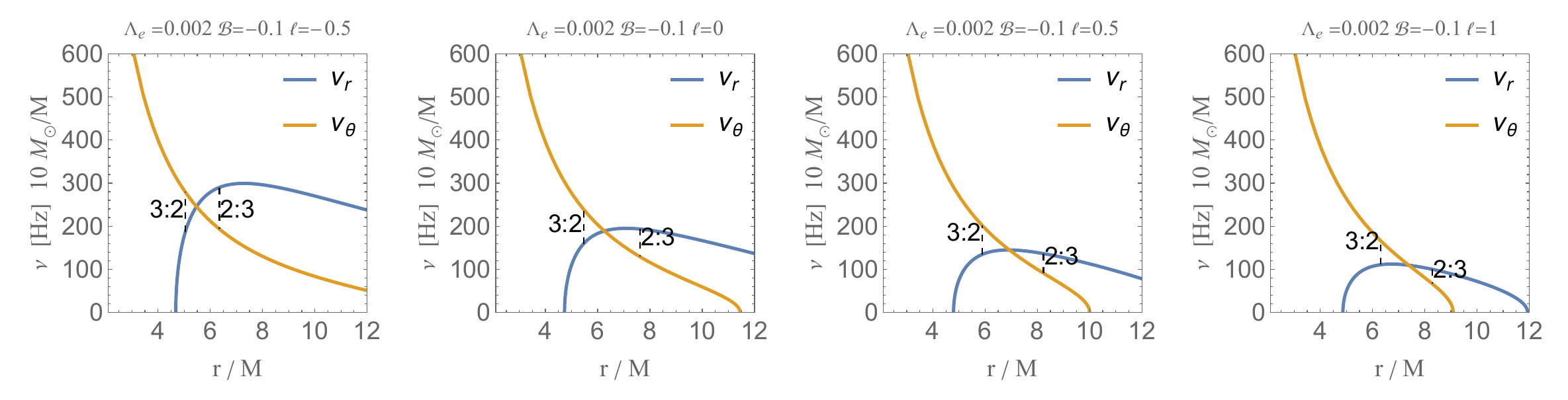}
    \caption{The epicyclic frequencies of the charged particle around the Schwarzschild-like black hole with the positive cosmological constant in the bumblebee gravity immersed in an external uniform magnetic field. Cosmological constant $\Lambda_{e}$,LSB parameter $\ell$ and  magnetic parameter $\mathcal{B}$ are chosen to be $\Lambda_{e}=0.002$,$\ell=-0.5,0,0.5,1$, $\mathcal{B}=\pm 0.1$ respectively for illustration. The black dashed line represents the position where the ratio of $\nu_\theta$ and $\nu_r$ is 3:2 and 2:3.}
    \label{fig4}
\end{figure*}

From epicyclic resonant approach as shown in \cite{Kolos:2015iva}, $\nu_{r}$ and $\nu_{\theta}$ can be directly identified as the $\nu_{U}$ and $\nu_{L}$ in observations. Based on the Eqs.(\ref{frequency of r direction}),(\ref{frequency of theta direction}),(\ref{frequency in astronomy}), we plot the relation between the epicyclic frequencies of the charged particles and the circular orbit radius in Fig. \ref{fig3}.  
We also plot the behavior of frequencies for the case with positive cosmological constant for comparison in Fig.\ref{fig4}. By taking into account the positive cosmological constant, $\nu_{\theta}$ can approach zero at finite radius which is different from the case $\Lambda_{e}=0$ which can also be seen in Eq.(\ref{fds}).

From Fig.\ref{fig3} and Fig.\ref{fig4}, it can be directly seen that there are both inner resonance radius and outer resonance radius with frequency ratios 
\begin{equation}
    \nu_{\theta}:\nu_{r}=3:2, \quad \nu_{\theta}:\nu_{r}=2:3. 
\end{equation}
respectively for all different choices of parameters. Thus the observed QPO can come either from orbit $r_{3:2}$ or orbit $r_{2:3}$. For the radiation coming from $r_{3:2}$, $\nu_{\theta}$ is identified as $\nu_{U}$ in observation and $\nu_{r}$ is identified as $\nu_{L}$, while for radiation coming from $r_{2:3}$, $\nu_{\theta}\equiv \nu_{L}$ and $\nu_{r} \equiv \nu_{U}$.  This feature is the same as the case for Schwarzschild black hole in uniform magnetic field \cite{Kolos:2015iva}. Since the frequencies $\nu_{r}$ and $\nu_{\theta}$ depend sensitively on the Lorentz violation factor $\ell$ and have close relationship to the QPO observation, it will be very interesting to constrain the Lorentz symmetry breaking factor $\ell$ by using the observational data of QPO.  Thus in the next section, we will use Markov Chain Monte Carlo (MCMC) algorithm to constrain the Lorentz violation factor. 

\section{Constraints on the Lorentz violating factor by using MCMC analysis}  \label{mcmc}
In this section, we would like to use the observational results to constrain the Lorentz violating factor $\ell$ in bumblebee model. We will only focus on the case without cosmological constant, since in this case, the black hole shadow observation can not distinguish the spacetime Eq.(\ref{metric of bumblebee}) from Schwarzschild black hole due to degeneracy of effective potential which makes the constraint using QPO more necessary.  

\begin{figure}[h!]
    \centering   \includegraphics[width=0.8\linewidth]{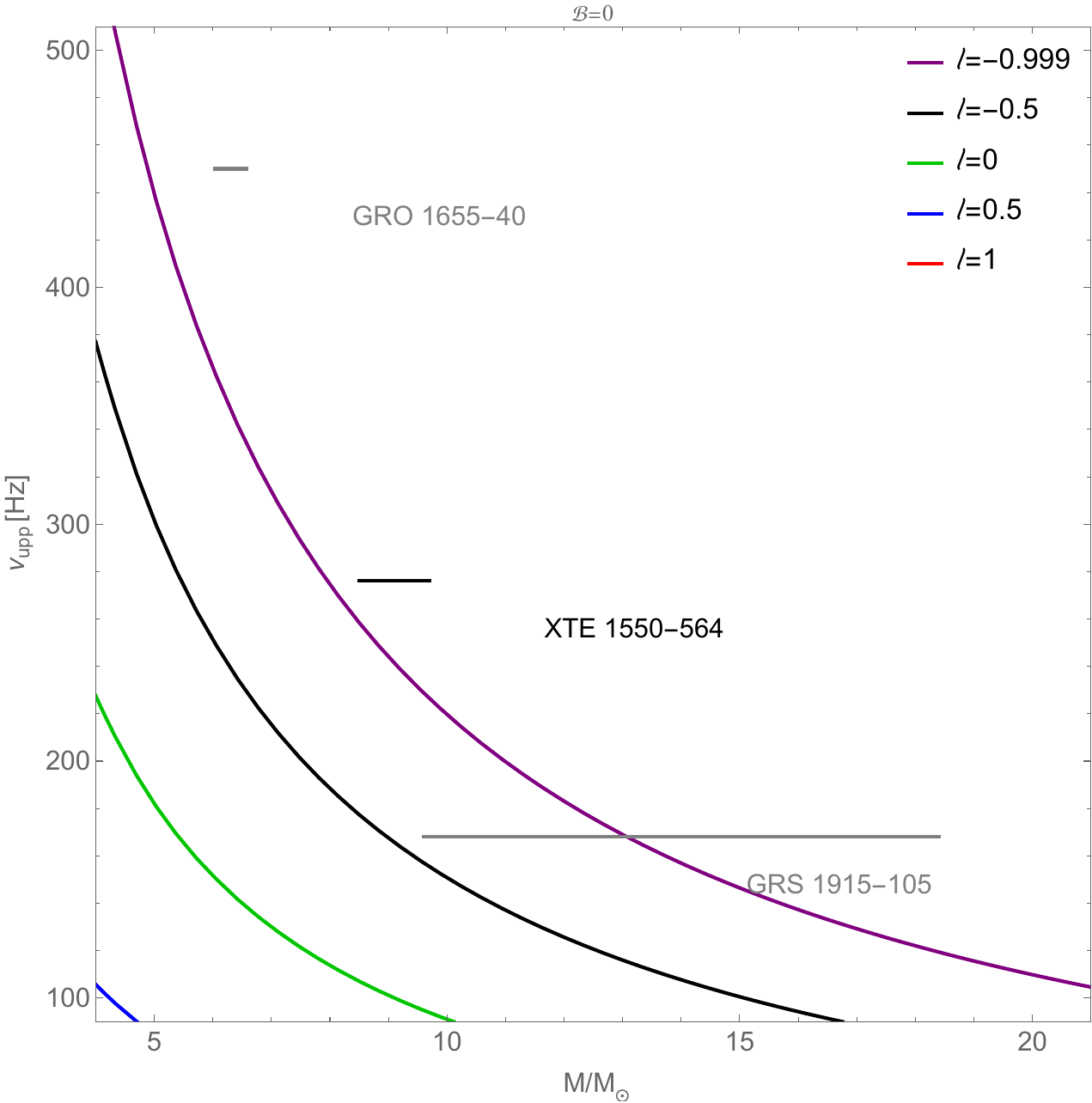}
    \caption{The relation between upper oscillation frequency $\nu_{U}=\nu_\theta$ and the black hole mass at the 3:2 inner resonance radius. In this plot, we set $\mathcal{B}=0$. Compared to the mass limits obtained from observations of the slowly rotating microquasars XTE J1550–564, we conclude that without magnetic field, the theoretical results do not match the observation.}\label{fig5}
\end{figure}

The observational results of three microquasars GRS 1915+105, XTE J1550-564 and GRO J1655-40 are respectively \cite{Kolos:2015iva}
\begin{equation}
\begin{aligned}
\textbf{GRO J1655-40}:\quad &\nu_U=450\pm 3\text{Hz},\nu_L=300^{+5}_{-5}\text{Hz},\\&M=6.3\pm 0.27 M_{\odot}
\end{aligned}\label{gro}
\end{equation}
\begin{equation}
\begin{aligned}
\textbf{XTE J1550-564}:\quad  &\nu_U=276\pm3\text{Hz}, \nu_L=184\pm5\text{Hz},\\& M=9.1 \pm 0.6 M_{\odot}
\end{aligned}\label{xte}
\end{equation}
\begin{equation}
\begin{aligned}
\textbf{GRS 1915+105}:\quad &\nu_U=168\pm3\text{Hz}. \nu_L=113\pm5\text{Hz}, \\& M=14 \pm 4.4 M_{\odot}
\end{aligned}\label{grs}
\end{equation}
However, below we will only focus on the results given by microquasars XTE J1550-564 as this source is slowly rotating which is close to our setup. We first choose the inner resonant radius $r_{3:2}$ as representative and plot the relation between upper frequencies $\nu_{\theta}$ and the mass of central black holes. As can be seen in Fig.\ref{fig5}, without magnetic field, the curve can not intersect the parameter region of XTE J1550-564 unless $\ell<-1$ which is theoretically unacceptable. 
Therefore, the theoretical results can not match the observation of XTE J1550-564 in the absence of magnetic field which is similar to Schwarzschild black hole \cite{Kolos:2015iva}. However, with the synergy of uniform magnetic field, 
the theoretical prediction of bumblebee black hole can meet the requirement of the observation in XTE J1550-564. We will investigate the cases for different resonance radius (3:2 radius or 2:3 radius) and different orientations of magnetic field in the following to find the best fit parameter values that can satisfy the observations .  

To obtain the best-fit values for the parameters, we employed \textit{emcee} for the MCMC analysis in the epiyclic resonance model. 

Based on Bayes' theorem, the posterior probability distribution of a model parameter $\Theta$ conditioned on observed data $\mathcal{D}$ can be mathematically formulated as
\begin{equation}
    \mathcal{P}(\Theta \mid \mathcal{D}) = \frac{P(\mathcal{D} \mid \Theta) P(\Theta)}{P(\mathcal{D})}
\end{equation}
where $\mathcal{P}(\Theta \mid \mathcal{D})$ represents the posterior probability which is updated belief about parameters after using the observational data, $P(\mathcal{D} \mid \Theta)$ represents the likelihood function of the observational data, $P(\Theta)$ is the prior probability of parameters and $P(\mathcal{D})$ is the normalization factor which ensures the probability integrate to 1. In our setup, the observational data $\mathcal{D}$ correspond to the quasi-periodic oscillation frequencies observed in each microquasars as shown in Eqs.(\ref{gro}),(\ref{xte}),(\ref{grs}), and $\Theta$ characterizes the the model parameters involved in the QPO phenomenon. 

We utilize the observational data of QPO in XTE J1550-564 during the MCMC fitting, thus the form of the likelihood function $\mathscr{L}$ is as follows:
\begin{equation}\log\mathscr{L}=\log\mathscr{L}_{\nu_{r}}+\log\mathscr{L}_\mathcal{\nu_{\theta}},\end{equation}
with
\begin{equation}\begin{gathered}
\log\mathscr{L}_{\nu_{r}}=-\frac{1}{2}\sum_i\frac{\left(\nu_{\nu_{r},\mathrm{obs}}^i-\nu_{\nu_{r},th}^i\right)^2}{\left(\sigma_{\nu_{r},\mathrm{obs}}^i\right)^2}, \\
\log{\mathscr{L}_\mathcal{\nu_{\theta}}}=-\frac{1}{2}\sum_i\frac{\left(\nu_{\mathcal{\nu_{\theta}},obs}^i-\nu_{\mathcal{\nu_{\theta}},th}^i\right)^2}{\left(\sigma_{\mathcal{\nu_{\theta}},obs}^i\right)^2}.
\end{gathered}\end{equation}
where the subscript ‘$obs$’ denotes the corresponding observed quantities while the subscript ‘$th$’ denotes the corresponding quantities calculated by the theoretical model.

For MCMC algorithm, more important thing is the choice of the prior distributions of the parameters. The choice of prior distribution of parameters $\{M,\mathcal{B},r_{0}\}$ can be given based on following reasons. The Lorentz violation is constrained to be vanishingly small in solar system by observations such as light bending \cite{Casana:2017jkc}. Going across to the strong gravity regime, even though the Lorentz violating factor may not be particularly small, it is still natural to expect that it will also not be excessively large in magnitude.\footnote{This expectation can also be corroborated by the constraints obtained from other theoretical settings.\cite{Yang:2023wtu,Junior:2024ety}} Thus in the presence of Lorentz violation factor $\ell$, its influence on other parameters such as magnetic parameter will not be significant. Based on the above physical reasons, we will use the following strategy to limit the prior distribution: we initially fit the parameters of magnetized Schwarzschild black hole ($\ell=0$) to the data by using MCMC which provides the estimation of black hole mass $M$, circular orbit radius $r_0$,and magnetic parameter $\mathcal{B}$. The resulting Gaussian distributions 
\begin{equation}\begin{gathered}
    P\left(\theta_{i}\right)\sim\exp\left[-\frac{1}{2}\left(\frac{\theta_{i}-\theta_{o,i}}{\sigma_{i}}\right)^{2}\right],\\\theta_{\mathrm{low},i}<\theta_{i}<\theta_{\mathrm{high},i},\theta_{i}=\left[\mathcal{B},r_0,M\right],
\end{gathered}\end{equation}
with mean and variance listed in Table.\ref{table1} can be assigned as the prior distribution for the magnetized bumblebee black hole. Moreover, as we have no knowledge about the Lorentz violating factor $\ell$, the prior distribution of $\ell$ is set to be a uniform distribution. 

\begin{table*}[ht!]
\centering
\caption{The prior distribution of parameters for the magnetized Schwarzschild-like bumblebee black hole}
\begin{tabular}{lcccc}

\midrule
\multirow{2}{*}{\textbf{Parameters}} & \multicolumn{4}{c}{\textbf{Resonance oscillation model of twin HF QPOs}} \\
\cmidrule(lr){2-5}
& \textbf{$\nu_{\theta}:\nu_{r}=3:2$, $B<0$} & \textbf{$\nu_{\theta}:\nu_{r}=2:3$, $B<0$} & \textbf{$\nu_{\theta}:\nu_{r}=3:2$, $B>0$} & \textbf{$\nu_{\theta}:\nu_{r}=2:3$, $B>0$} \\
& \textit{$\mu \pm \sigma$} & \textit{$\mu \pm \sigma$} & \textit{$\mu \pm \sigma$} & \textit{$\mu \pm \sigma$} \\
\midrule
$M\ (M_\odot)$ & $9.1607\ 0.34175$       & $9.3363\ 0.30001$         & $9.096\ 0.34120$        & $9.1269\ 0.32135$       \\
$\mathcal{B}$       & $-0.0834\ 0.01340$ & $-0.2040\ 0.07050$     & $0.0475\ 0.00375$  & $0.0530\ 0.00255$ \\
$r_{0}/M$           & $5.4731\ 0.13605$       & $7.1293\ 0.17205$         & $5.4983\ 0.13695$        & $7.1977\ 0.17200$        \\
$\ell$           & Uniform[-1,1]        & Uniform[-1,1]         & Uniform[-1,1]         & Uniform[-1,1]        \\
\bottomrule
\label{table1}
\end{tabular}
\end{table*}

\begin{figure*}[ht!]
  \centering
  \begin{minipage}[t]{0.48\textwidth} 
    \centering
    \includegraphics[width=1.2\textwidth, height=9cm, keepaspectratio]{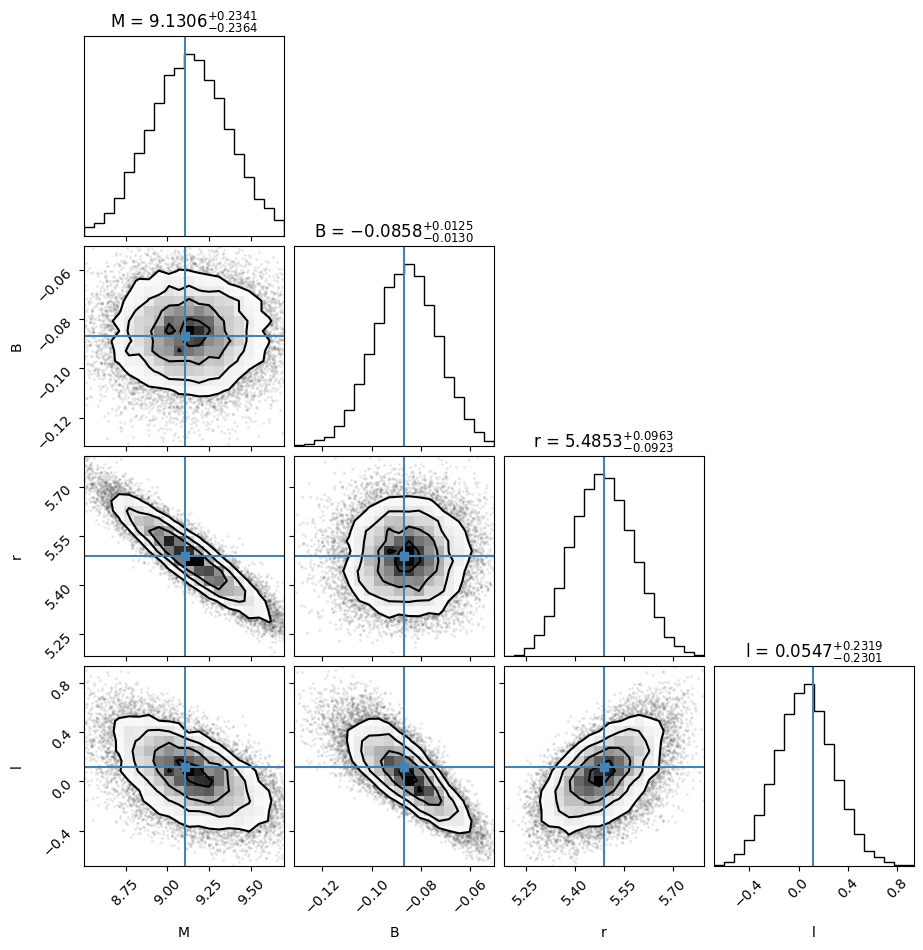} 
    \textbf{(a) Frequency ratio $\nu_{\theta}:\nu_{r}=3:2$ and downward magnetic field }
    \label{fig:img1}
  \end{minipage}%
  \hfill 
  \begin{minipage}[t]{0.48\textwidth}%
    \centering
    \includegraphics[width=1.2\textwidth, height=9cm, keepaspectratio]{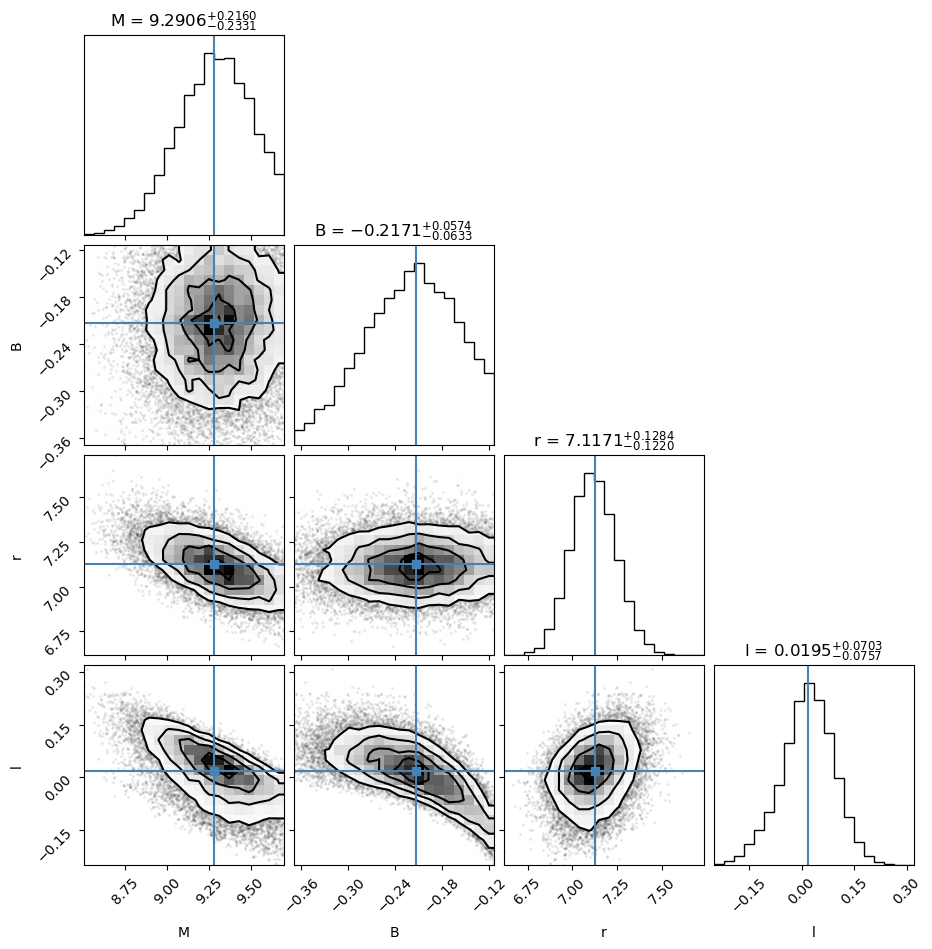}
    \textbf{(b) Frequency ratio $\nu_{\theta}:\nu_{r}=2:3$ and downward magnetic field}
    \label{fig:img2}
  \end{minipage}

  \vspace{0.5cm} 

  \begin{minipage}[t]{0.48\textwidth}%
    \centering
    \includegraphics[width=1.2\textwidth, height=9cm, keepaspectratio]{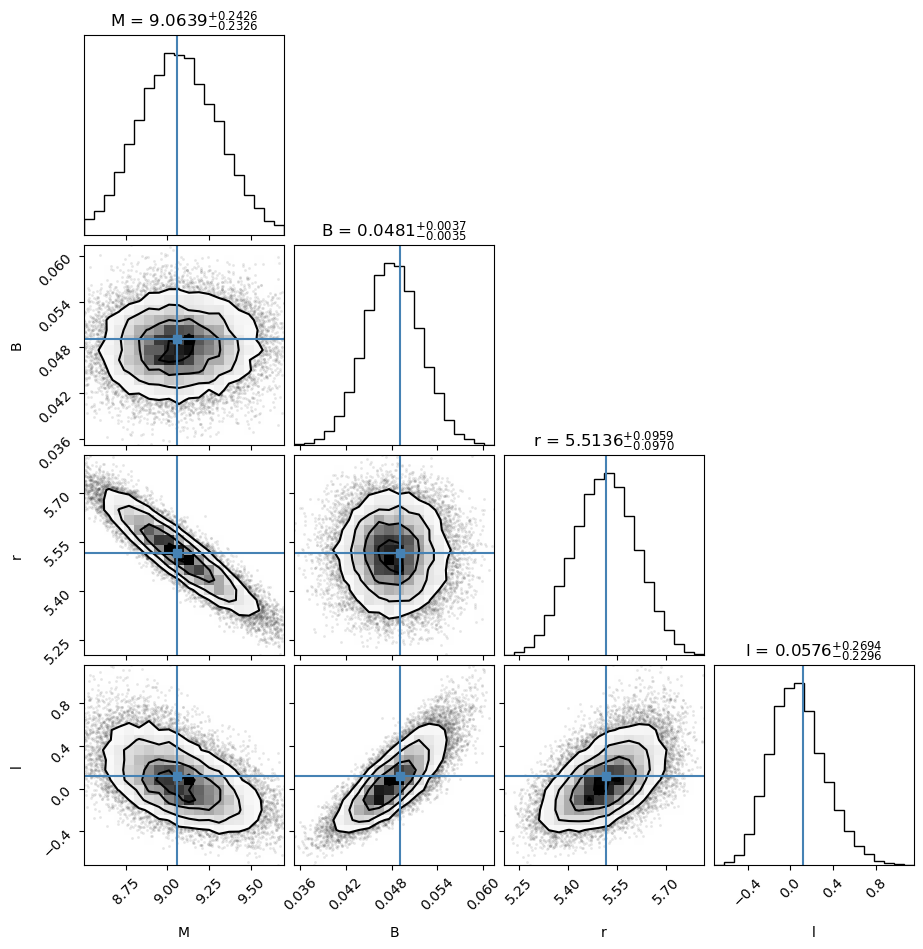}
    \textbf{(c) Frequency ratio $\nu_{\theta}:\nu_{r}=3:2$ and upwards magnetic field}
    \label{fig:img3}
  \end{minipage}%
  \hfill
  \begin{minipage}[t]{0.48\textwidth}%
    \centering
    \includegraphics[width=1.2\textwidth, height=9cm, keepaspectratio]{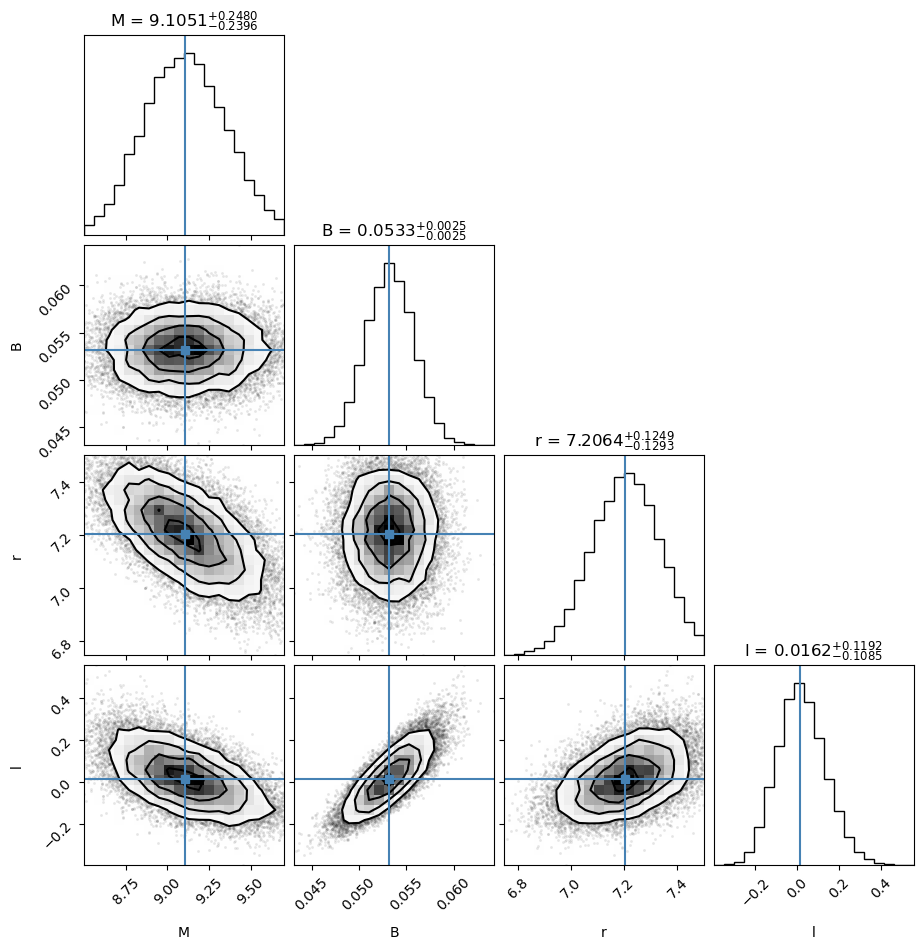}
    \textbf{(d) Frequency ratio $\nu_{\theta}:\nu_{r}=2:3$ and upwards magnetic field}
    \label{fig:img4}
  \end{minipage}

  \caption{Final constraints on the parameters of the magnetized bumblebee black hole by using QPO observational data of micro-quasar XTE J1550-564. } 
  \label{fig:grid}
\end{figure*}

Subsequently, based on the previously determined prior distribution, we employ the MCMC techniques to constrain the parameter values of $\{M,\mathcal{B},r_{0},\ell \}$ for the magnetized bumblebee black hole with Lorentz violation. As both inner resonance radius ($r_{3:2}$) and outer resonance radius ($r_{2:3}$) can produce the QPO signal and the orientation of magnetic field remains obscure, we investigate all the four cases for different choice of  resonance radius and magnetic field orientations. We randomly generate 20000 points for each parameter to obtain the corresponding best-fit values. The corner diagram has been plotted in Fig.\ref{fig:grid} where we use the blue dot to mark the most probable value of each parameter. The final parameter values which best fits the observational data are presented in Table.\ref{table2}. 

Based on the results obtained by MCMC, for the four cases shown in Fig.\ref{fig:grid}, the Lorentz violating factor $\ell$ is constrained (at 68\% confidence level) to be $[-0.1754,0.2866]$, $[-0.0562,0.0898]$, $[-0.172,0.327]$, $[-0.0923,0.1354]$ respectively. Apart from Schwarzschild black hole in Einstein gravity, the data can also support Schwarzschild-like bumblebee black holes with Lorentz violation. MCMC analysis of QPO observation can help to constrain the  parameter regions of $\ell$ which is not possible by using black hole shadow observation because of the degeneracy as we discussed in Sec.\ref{motion}. Therefore, the results we reported in this work will be crucial for examining the theory. Stricter constraints can only be given when we know more detailed information about the magnetic field near the black hole by other observations, which will further constrain the Lorentz violation and finally determine whether Lorentz violation appears in strong gravity regime.

\begin{table*}[ht!]
\centering
\caption{The best-fit values of parameters for the magnetized bumblebee black hole by using the resonance oscillation model}
\begin{tabular}{lcccc}

\midrule
\multirow{2}{*}{\textbf{Parameters}} & \multicolumn{4}{c}{\textbf{Epicyclic resonance oscillation model of twin HF QPOs}} \\
\cmidrule(lr){2-5}
& \textbf{$\nu_{\theta}:\nu_{r}=3:2$, $B<0$ } & \textbf{$\nu_{\theta}:\nu_{r}=2:3$, $B<0$} & \textbf{$\nu_{\theta}:\nu_{r}=3:2$, $B>0$} & \textbf{$\nu_{\theta}:\nu_{r}=2:3$, $B>0$} \\
\midrule
$M\ (M_\odot)$ & $9.1306_{-0.2364}^{+0.2341}$       & $9.2906_{-0.2331}^{+0.2160}$         & $9.0639_{-0.2326}^{+0.2426}$        & $9.1051_{-0.2396}^{+0.2480}$       \\
$\mathcal{B}$       & $-0.0858_{-0.0130}^{+0.0125}$ & $-0.2171_{-0.0633}^{+0.0574} $     & $0.0481_{-0.0035}^{+0.0037} $  & $0.0533_{-0.0025}^{+0.0025} $ \\
$r_{0}/M$           & $5.4853_{-0.0923}^{+0.0963}$       & $7.1171_{-0.1220}^{+0.1284} $         & $5.5136_{-0.0970}^{+0.0959} $        & $7.2064_{-0.1293}^{+0.1249} $        \\
$\ell$           & $0.0547_{-0.2301}^{+0.2319}$         & $0.0195_{-0.0757}^{+0.0703}$         & $0.0576_{-0.2296}^{+0.2694}$        & $0.0162_{-0.1085}^{+0.1192}$        \\
\bottomrule
\label{table2}
\end{tabular}
\end{table*}

\section{Conclusion and discussion}\label{conclusion}
In this work, we have explored the harmonic oscillation dynamics of charged particles for Schwarzschild-like bumblebee black hole immersed in uniform magnetic field. We investigate the effect of Lorentz violating factor $\ell$ on the particle motion. After that, by using the twin peak HF QPO data of microquasar XTE J1550-564 and relating it to the epicyclic oscillation frequencies of charged particles, we employ the Markov Chain Monte Carlo algorithm to constrain the Lorentz violating factor. The constrains on the Lorentz violating factor $\ell$ are summarized in Table \ref{table2}. Our work shows that the degeneracy between Schwarzschild and Schwarzschild-like bumblebee black holes in black hole shadow (circular orbits) can be circumvented through other dynamical probes related to observation such as epi-cyclic oscillation frequencies, which sensitively depend on the Lorentz violating factor. The constraints we obtained in this work will be crucial for further searching for the imprint of Lorentz symmetry violation in our universe.  

For future directions, from the theoretical side, the above analysis should be generalized to rotating magnetized black hole case. The central obstacle is that there are still some debates on the correct form of rotating bumblebee black hole.  While rotating bumblebee black hole is derived from static spherical symmetric black hole by using Newman-Janis algorithm \cite{Ding:2019mal}, there are still some discussions \cite{Maluf:2022knd} on if the algorithm works.  From the observational sides, since our $1\sigma$ confidence interval of $l$ contains $l=0$ case, it is too soon to get any concrete conclusion from observation on whether there is Lorentz symmetry breaking in strong gravity regime. However, further observational data on the magnetic field around black hole will lead to more strict parameter estimation which is very important for searching for the Lorentz symmetry breaking. 

\section*{Acknowledgements}
This work is supported by the National Natural Science Foundation of China (NSFC) under Grant Nos.12405066, 12175105. YSA is also supported by the Natural Science Foundation of Jiangsu Province under Grant No.BK20241376 and Fundamental Research Funds for the Central Universities. 








\bibliographystyle{elsarticle-num} 
\bibliography{reference}

\end{document}